\documentclass[12pt]{iopart}

\usepackage{epsfig}
\usepackage{scalefnt}
\begin{document}

\title[Interpreting $f_0(600)$ and $a_0(980)$ as $\bar q q$ states
from a $U(3) \times U(3)$ Sigma Model with (Axial-)Vectors]{Interpreting \boldmath $f_0(600)$ and \boldmath $a_0(980)$ as
\boldmath $\bar q q$ states
from an \boldmath $N_f=3$ Sigma Model with (Axial-)Vectors}{Nordic Conference on Nuclear Physics, Stockholm, 13-17 June 2011}

\author{Denis Parganlija}

\address{Institute for Theoretical Physics, Goethe University,
Max-von-Laue-Str.\ 1, D--60438 Frankfurt am Main, Germany        
}
\ead{parganlija@th.physik.uni-frankfurt.de}
\begin{abstract}
We address the question whether it is possible to interpret the low-lying scalar mesons
$f_0(600)$ and $a_0(980)$ as $\bar q q$ states within a $U(3) \times U(3)$ Linear Sigma Model containing
vector and axial-vector degrees of freedom.
\end{abstract}


\section{Introduction}

The structure of scalar mesons has been a matter of debate for many decades. Experimental data \cite{PDG} suggest the existence
of at least five $IJ^{PC}=00^{++}$ states in the region up to 1.75 GeV: $f_0(600)$, $f_0(980)$, $f_0(1370)$, $f_0(1500)$ and $f_0(1710)$.
In the scalar sector, the existence of the scalar $K^{\star}_0(1430)$ state has been confirmed, unlike the existence of the
low-lying scalar kaon $K^{\star}_0(800)$ (or $\kappa$).  
Kaons
and other mesons containing strange quarks are expected to play an important
role in vacuum phenomenology and also in the restoration of the $%
U(N_{f})_{L}\times U(N_{f})_{R}$ chiral symmetry \cite{Reference1}, a
feature of the Quantum Chromodynamics (QCD) broken in vacuum spontaneously 
\cite{Goldstone} by the quark condensate and explicitly by non-vanishing
quark masses ($N_{f}$: number of quark flavours). 
\\
Meson
phenomenology has been considered in
various $\sigma$ model approaches (see \cite{Reference1} and references
therein). In this paper, we present an Extended Linear Sigma Model (eLSM 
\cite{LesHouches,Krakow,Madrid}) containing
scalar, pseudoscalar, vector and axial-vector mesons both in the non-strange
and strange sectors. We have addressed the issue of structure of
scalar mesons $f_0(1370)$ and $f_0(1710)$ in \cite{LesHouches}; the conclusion was that one could find a reasonable global
fit of scalar, pseudoscalar, vector and axial-vector masses within a single model assuming
that $a_0(1450)$ and $K^{\star}_0(1430)$ are $\bar q q$ states. A consequence of the fit was the
statement that $f_0(1370)$ and $f_0(1710)$ correspond to scalar $\bar q q$ states. 
However,
\cite{LesHouches} did not address a different possibility: whether such a fit
could also be found assuming that $a_0(980)$ and $\kappa$, rather than $a_0(1450)$ and $K^{\star}_0(1430)$,
are $\bar q q$ states. If a reasonable fit exists in this case,
then the results of \cite{LesHouches} are inconclusive; however, if the opposite is true, then 
the suggestion is confirmed that scalars above 1 GeV are favoured to be $\bar q q$ states.
\\
The paper is organised as follows. In Sec.\ 2 we present the
Lagrangian, the results are discussed in Sec.\ 3 and in Sec.\ 4 we provide a
summary and outlook of further work.

\section{The Model}

The Lagrangian of the Extended Linear Sigma Model with $U(3)_{L}\times
U(3)_{R}$ symmetry reads \cite{Reference1,Krakow,Madrid}: 
\begin{eqnarray}
\mbox{\fontsize{10}{9}\selectfont $ \lefteqn{\mathcal{L}=\mathrm{Tr}[(D^{\mu }\Phi )^{\dagger }(D^{\mu }\Phi)]-m_{0}^{2}\mathrm{Tr}(\Phi ^{\dagger }\Phi )-\lambda _{1}[\mathrm{Tr}(\Phi^{\dagger }\Phi )]^{2}-\lambda _{2}\mathrm{Tr}(\Phi ^{\dagger }\Phi )^{2}} $} \nonumber \\
& & \mbox{\fontsize{10}{9}\selectfont $ - \, \frac{1}{4}\mathrm{Tr}[(L^{\mu \nu })^{2}+(R^{\mu \nu })^{2}]+\mathrm{Tr}\left[ \left( \frac{m_{1}^{2}}{2}+\Delta \right) (L^{\mu })^{2}+(R^{\mu})^{2}\right] +\mathrm{Tr}[H(\Phi +\Phi ^{\dagger })] $} \nonumber \\
&&  \mbox{\fontsize{10}{9}\selectfont $ + \, c_{1}(\det\Phi-\det\Phi^{\dagger})^{2}+i\frac{g_{2}}{2}(\mathrm{Tr}\{L_{\mu \nu }[L^{\mu },L^{\nu }]\}+\mathrm{Tr}\{R_{\mu \nu }[R^{\mu },R^{\nu }]\}) $}   \nonumber \\
&&  \mbox{\fontsize{10}{9}\selectfont $ +\, \frac{h_{1}}{2}\mathrm{Tr}(\Phi ^{\dagger }\Phi )\mathrm{Tr}[(L^{\mu})^{2}+(R^{\mu })^{2}]+h_{2}\mathrm{Tr}[(\Phi R^{\mu })^{2}+(L^{\mu }\Phi)^{2}]+2h_{3}\mathrm{Tr}(\Phi R_{\mu }\Phi ^{\dagger }L^{\mu }) $} 
\label{Lagrangian}
\end{eqnarray}
where
\begin{equation}
\scalefont{0.81}\Phi =\frac{1}{\sqrt{2}}\left( 
\begin{array}{ccc}
\frac{(\sigma _{N}+a_{0}^{0})+i(\eta _{N}+\pi ^{0})}{\sqrt{2}} & 
a_{0}^{+}+i\pi ^{+} & K_{S}^{+}+iK^{+} \\ 
a_{0}^{-}+i\pi ^{-} & \frac{(\sigma _{N}-a_{0}^{0})+i(\eta _{N}-\pi ^{0})}{%
\sqrt{2}} & K_{S}^{0}+iK^{0} \\ 
K_{S}^{-}+iK^{-} & {\bar{K}_{S}^{0}}+i{\bar{K}^{0}} & \sigma _{S}+i\eta _{S}%
\end{array}%
\right)  \normalsize \label{Phi}
\end{equation}%
%
%
%
%
%
is a matrix containing the scalar and pseudoscalar degrees of freedom, $%
L^{\mu }=V^{\mu }+A^{\mu }$ and $R^{\mu }=V^{\mu }-A^{\mu }$ are,
respectively, the left-handed and the right-handed matrices containing
vector and axial-vector degrees of freedom with 
\begin{equation}
\scalefont{0.81}V^{\mu }=\frac{1}{\sqrt{2}}\left( 
\begin{array}{ccc}
\frac{\omega _{N}+\rho ^{0}}{\sqrt{2}} & \rho ^{+} & K^{\star +} \\ 
\rho ^{-} & \frac{\omega _{N}-\rho ^{0}}{\sqrt{2}} & K^{\star 0} \\ 
K^{\star -} & {\bar{K}}^{\star 0} & \omega _{S}%
\end{array}%
\right) ^{\mu }{\normalsize ,}\;\scalefont{0.81}A^{\mu }=\frac{1}{\sqrt{2}}%
\left( 
\begin{array}{ccc}
\frac{f_{1N}+a_{1}^{0}}{\sqrt{2}} & a_{1}^{+} & K_{1}^{+} \\ 
a_{1}^{-} & \frac{f_{1N}-a_{1}^{0}}{\sqrt{2}} & K_{1}^{0} \\ 
K_{1}^{-} & {\bar{K}}_{1}^{0} & f_{1S}%
\end{array}%
\right) ^{\scalefont{0.81}\mu } \normalsize \label{LR}
\end{equation}
and $\Delta =\mathrm{diag}(\delta _{N},\delta _{N},\delta _{S})$ describes
explicit breaking of the chiral symmetry in the (axial-)vector channel. 
The explicit symmetry breaking in the (pseudo)scalar sector is
described by Tr$[H(\Phi +\Phi ^{\dagger })]$ with $H=1/2\,\mathrm{%
diag}(h_{0N},h_{0N},\sqrt{2}h_{0S})$, $h_{0N}=const.$, $h_{0S}=const.$ Also, 
$D^{\mu }\Phi =\partial ^{\mu }\Phi -ig_{1}(L^{\mu }\Phi -\Phi R^{\mu })$ is the covariant derivative; 
$L^{\mu \nu }=\partial ^{\mu }L^{\nu }-\partial ^{\nu }L^{\mu }$, $%
R^{\mu \nu }=\partial ^{\mu }R^{\nu }-\partial
^{\nu }R^{\mu }$ are, respectively, the
left-handed and right-handed field strength tensors and the term $%
c_{1}(\det\Phi-\det\Phi^{\dagger})^{2}$
describes the $U(1)_{A}$ anomaly \cite{Klempt}.
\\
We assign the fields $\vec{\pi}$ and $\eta _{N}$
to the pion and the $SU(2)$ counterpart of the $\eta $ meson, $\eta
_{N}\equiv (\bar{u}u+\bar{d}d)/\sqrt{2}$. The fields $\omega
_{N}^{\mu }$, $\vec{\rho}^{\mu }$, $f_{1N}^{\mu }$ and $\vec{a}_{1}^{\mu }$
are assigned to the $\omega (782)$, $\rho (770)$, $f_{1}(1285)$ and $%
a_{1}(1260)$ mesons, respectively \cite{Paper1}. We also
assign the $K$ fields to the kaons; $\eta _{S}$ is the strange contribution
to the $\eta $ and $\eta ^{\prime }$ fields and the $\omega _{S}^{\mu } $, $%
f_{1S}^{\mu }$, $K^{\star \mu }$ and $K_{1}^{\mu }$ fields correspond to the 
$\varphi (1020)$, $f_{1}(1420)$, $K^{\star }(892)$ and $K_{1}(1270)$ mesons,
respectively.
\\
The assignment of the scalar states in our model to physical resonances is
ambiguous. In this work, we assign the $\vec{a}_{0}
$ field to $a_{0}(980)$ and $K_{S}$ to the physical $%
K_{0}^{\star }(800)$ state. We thus presuppose that these two
states below 1 GeV are $\bar{q}q$ states (as all the fields present in
our model are $\bar{q}q$\ states \cite{Paper1}) and describe in the following section whether 
a global fit of masses can be found under this assumption.
\\
The Lagrangian (\ref{Lagrangian}) also contains two $IJ^{PC}=00^{++}$
states, $\sigma _{N}$ (pure $\bar n n$ state with $n \in \{ u,d$ quarks$\}$) and $\sigma
_{S}$ (pure strange state, $\sigma _{S}\equiv \bar{s}s$). The states $\sigma _{N}$ and $\sigma
_{S}$ mix and two new states emerge: $\sigma
_{1}$ (predominantly non-strange) and $\sigma _{2}$ (predominantly strange) \cite{LesHouches}. 
We describe the assignment of $\sigma _{1,2}$ to physical states in the next section.
\\
In order to implement spontaneous symmetry breaking in the model, we shift $%
\sigma _{N}$ and $\sigma _{S}$ by their respective vacuum expectation values 
$\phi _{N}$\ and $\phi _{S}$ (see \cite{LesHouches,Madrid} for details). 
\\
The Lagrangian (\ref{Lagrangian}) contains 14 parameters: $\lambda _{1}$, $%
\lambda _{2}$, $c_{1}$, $h_{0N}$, $h_{0S}$, $h_{1}$, $h_{2}$, $h_{3}$, $m_{0}^{2}$, $%
g_{1}$, $g_{2}$, $m_{1}$, $\delta _{N}$, $\delta _{S}$. The parameter $g_{2}$
is determined from the decay width $\rho \rightarrow \pi \pi $ \cite{Paper1}%
; we set $h_{1}=0$ as this parameter is large-$N_{c}$ suppressed \cite%
{Paper1} and also $\delta _{N}=0$ because the explicit symmetry breaking is
small in the non-strange sector. All other parameters are calculated from a
global fit of masses including $m_{\pi }$, $m_{K}$, $m_{\eta }$, $m_{\eta
^{\prime }}$, $m_{\rho }$, $m_{K^{\star }}$, $m_{\omega _{S}\equiv \varphi
(1020)}$, $m_{f_{1S}\equiv f_{1S}(1420)}$, $m_{a_{1}}$, $m_{K_{1}\equiv
K_{1}(1270)}$, $m_{a_{0}\equiv a_{0}(980)}$ and $m_{K_{S}\equiv
\kappa}$. Note, however, that the mass terms
from the Lagrangian (\ref{Lagrangian})\ used in the fit allow only for the
linear combination $m_{0}^{2}+\lambda _{1}(\phi _{N}^{2}+\phi _{S}^{2})$
rather than the parameters $m_{0}^{2}$ and $\lambda _{1}$ by themselves to
be determined.

\section{The Global Fit and Two-Pion Decay of the Sigma Mesons}
Results from our best fit are shown in Table \ref{Table1}. We note that $m_{\kappa}$, $m_{a_{1}}$, $m_{K_{1}}$, $m_{\omega_{S}}$ and $m_{f_{1S}}$ deviate substantially
from their respective experimental values \cite{PDG}. 
This is in particular a problem for the rather sharp resonances $\omega_S \equiv \varphi(1020)$ and $f_{1}(1420) \equiv f_{1S}$.
\begin{table}
\caption{\label{Table1}Masses from our global fit.} 
\begin{indented}
\item[]\begin{tabular}{@{}*{7}{l}} 
\br                              
$  \mbox{Mass} $&$m_{\pi }$ &$m_{K}$&$m_{\eta }$ &$m_{\eta ^{\prime }}$&$m_{\rho
}$ &$m_{K^{\star }}$ \cr 
\mr
PDG Value (MeV) \cite{PDG} & 139.57 & 493.68 & 547.85 & 957.78 & 775.49 & 
891.66\cr 
\mr
Our Value (MeV) & 138.04 & 490.84 & 517.13 & 957.78 & 775.49 & 832.53\cr
\mr
Mass & $m_{\varphi }$ & $m_{f_{1S}}$ & $m_{a_{1}}$ & $m_{K_{1}}$ & $%
_{a_{0}} $ & $m_{K_{S}}$\cr 
\mr
PDG Value (MeV) \cite{PDG} & 1019.5 & 1426.4 & 1230 & 1272 & 980 & 676 \cr
\mr
Our Value (MeV) & 870.35 & 1643.4 & 1395.5 & 1520 & 978.27 & 1128.7 \cr 
\br
\end{tabular}
\end{indented}
\end{table}
In the scalar sector, if we set
$m_{\sigma_{1}}=705$ MeV and $m_{\sigma_{2}}=1200$ MeV then we obtain $\Gamma_{\sigma_{1}\rightarrow\pi\pi}=305$ MeV and $\Gamma_{\sigma
_{2}\rightarrow\pi\pi}=207$ MeV as well as $\Gamma_{\sigma
_{1}\rightarrow KK}=0$ and $\Gamma_{\sigma_{2}\rightarrow KK}=240$ MeV. 
These results correspond very well to experiment \cite{PDG,buggf0}. We thus assign $\sigma_{1}$ to $f_{0}(600)$
and $\sigma_{2}$ to $f_{0}(1370)$. Consequently, $f_{0}(600)$ is interpreted
as a predominantly non-strange $\bar{q}q$ state while $f_{0}(1370)$ is
interpreted as a predominantly $\bar{s}s$ state. The results also suggest,
however, that $f_{0}(1370)$ should predominantly decay into kaons (as
$\Gamma_{\sigma_{2}\rightarrow KK}/\Gamma_{\sigma_{2}\rightarrow\pi\pi}=1.15$)
-- not surprising for a $\bar{s}s$ state but clearly at odds with
experimental data \cite{PDG}.
\\
Additionally, the phenomenology in the vector and axial-vector channels
is not well described. The decay width $\Gamma
_{a_{1}(1260)\rightarrow\rho\pi}$ depends on parameter $g_2$, fixed via $\Gamma_{\rho\rightarrow\pi\pi}$ \cite{Paper1}.
A calculation of the decay width $\Gamma
_{a_{1}(1260)\rightarrow\rho\pi}$ then yields values of more than 10 GeV if we set $\Gamma_{\rho\rightarrow\pi\pi} = 149.1$ MeV 
(as suggested by the PDG \cite{PDG}).
Alternatively, if one
forces $\Gamma_{a_{1}(1260)\rightarrow\rho\pi}<600$ MeV to comply with the
data, then $\Gamma_{\rho\rightarrow\pi\pi}<38$ MeV is obtained -- a
value that is approximately $100$ MeV less than the experimental result.
\\
Then the fit results, and thus the assumption of
scalar $\bar q q$ states below 1 GeV, are problematic.

\section{Summary and Outlook}

We have presented a $U(3)_{L}\times U(3)_{R}$ Linear Sigma Model with
(axial-)vector mesons. A global fit of all masses (except the sigma masses) has been performed to determine model parameters.
The fit included masses of scalar states $\vec a_0$ and $K_{S}$, assigned respectively to $a_0(980)$ and $\kappa$.
The fit does not yield particulary good mass values: several of the masses considered deviate for more than 100 MeV from
the corresponding PDG value (see Table \ref{Table1}). We find very good correspondence of 
$\Gamma_{\sigma_{1,2}\rightarrow\pi\pi}$ and $\Gamma_{\sigma_{1,2}\rightarrow KK}$ to the experiment if one considers
$m_{\sigma_{1}}=705$ MeV and $m_{\sigma_{2}}=1200$ MeV. We thus assign $\sigma_1$ to $f_0(600)$ and $\sigma_2$ to $f_0(1370)$.
This implies that $f_0(1370)$ is predominantly a $\bar s s$ state.
Consequently, our results suggest that $f_0(1370)$ decays predominantly into kaons -- at odds with experiment \cite{PDG}.
Additionally, we cannot accomodate a correct (axial-)vector phenomenology into the fit: either $a_1(1260)$ is too broad ($\simeq 10$ GeV)
or the $\rho$ meson is too narrow ($\simeq 40$ MeV). We thus conclude that the fit of \cite{LesHouches}, where the scalar 
states were assumed to be above (rather than, as in this work, below) 1 GeV, described meson phenomenology 
better than the fit presented here and that this work confirmes 
the conclusion of \cite{LesHouches} that scalar $\bar q q$ states are favoured to have mass above 1 GeV.
\\
{\bf Acknowledgment.} I am grateful to Francesco Giacosa, Dirk Rischke, P\'{e}ter Kov\'{a}cs and Gy\"{o}rgy Wolf for valuable discussions regarding my work.

\end{document}